  \documentstyle[multicol,aps]{revtex}

\draft
\title{Fluctuation-dissipation theorem and flux noise \\
in overdamped Josephson junction arrays}
\author{S. E. Korshunov}
\address{L. D. Landau Institute for Theoretical Physics,
         Kosygina 2, 117940 Moscow, Russia}
\date{March 26, 2002}

\begin{document}
\maketitle

\begin{abstract}
The form of the fluctuation-dissipation theorem
for a resistively shunted Josephson junction array
is derived with the help of the method
which explicitely takes into account screening effects.
This result is used to express the flux noise power spectrum
in terms of frequency dependent sheet impedance of the array.
The relation between noise amplitude and parameters of the detection
coil is analysed for the simplest case of a single-loop coil.
\end{abstract}

  \begin{multicols}{2}
\section{Introduction}

The two basic experimental methods used for contactless investigation
of finite frequency properties of two-dimensional
superconducting systems (such as thin films
\cite{FH,Calame,Rogers,Festin,Ferrari}, Josephson junction arrays
\cite{Lerch,Shaw,Candi,ML} and wire networks \cite{JGRLM,Meyer}) are
two-coil mutual-inductance technique
\cite{FH,Calame,Rogers,Festin,ML,JGRLM,Meyer} and flux noise
power spectrum analysis \cite{Rogers,Festin,Ferrari,Lerch,Shaw,Candi}.
The first of them is based on measurement
of the voltage induced in the detection coil
by the currents flowing in the sample under the action of ac
electric field produced by the current in the other (driving) coil. For
the given geometry of the coils the measured signal can be used to
extract \cite{JGRLM} the complex frequency dependent sheet impedance
$Z_\Box(\omega)$ of the sample on the assumption that for wave-lengths
larger than the characteristic dimensions of the detection coil
$Z_\Box(\omega)$ is not wave-length dependent.

In the case of the flux noise spectrum analysis
the approaches to interpretation of the experimental data
are much more varied.
The theoretical predictions of the flux noise spectrum
used for comparison with experimental data
are found by relating it with \cite{Shaw,Timm}
or (no less often) by replacing it by \cite{Rogers,Houlrik,WF}
a correlation function describing the vortex distribution
and, naturally, turn out to be dependent on the particular choice of
assumptions concerning the form of this distribution.
Numerical simulations also demonstrate a clear tendency towards
studying the vortex number noise \cite{Houlrik,Jose,HS}
rather then the flux noise.
The only attempt to achieve a description of the flux
noise power spectrum in terms of the sheet impedance of the sample
taking into account the actual geometry of the
detection coil has been undertaken by Kim and Minnhagen
\cite{KM}. However, this calculation is also based on expressing all
quantities in terms of the vortex gas correlation functions
and, therefore, can be trusted only in a limited range of
parameters.

In the present work we argue that in the case of
a resistively shunted Josephson junction array
the general expression for the flux noise spectrum can be found
without artificial decomposition of all fluctuations into  the vortex
part (which is usually assumed to be responsible for the flux noise) and
the remaining so-called "spin-wave" part (which is traditionally
neglected). Although in
semi-phenomenological treatment \cite{Timm,Houlrik,WF,Minnh} of
two-dimensional superconductors such decomposition seems to be
inevitable, the case of an overdamped Josephson junction array
allows for application of a more universal approach to calculation of
the flux noise power spectrum.
It is based on the direct relation (discussed in Sec. 2)
of the flux noise with current correlations, which, on the other hand,
can be expressed in terms of the complex frequency dependent sheet
impedance with the help of the fluctuation-dissipation theorem.

The additional advantage of such approach is that it allows to include
into consideration in systematic way the mutual influence between
magnetic field fluctuations and current fluctuations (the screening
effects), 
which insofar has been neglected in theoretical works
\cite{Timm,Houlrik,WF,Jose,HS,KM} devoted to flux noise spectrum analysis.
The form of the Hamiltonian, which should be used for the description
of a resistively shunted array in presence of self-induced magnetic
fields, is discussed in Sec. 3,
and the corresponding dynamic equations in Sec. 4.

The explicit form of the fluctuation-dissipation theorem for
resistively shunted Josephson junction array is derived in Sec. 5. It
shows that the current correlations in the array are determined by the
response of the current to the { external} electric field and not
directly by the sheet impedance of the array (which is defined as a
response to the {\em total} electric field)
The nature of the expression for currents correlation function,
related with the peculiarities of the two-dimensional geometry,
allows to expect the same expression to be applicable for arbitrary
two-dimensional systems in which capacitive effects can be neglected.

Our main result, the relation between the flux noise power spectrum and
the frequency dependent sheet impedance of a two-dimensional
superconductor, is presented in Sec. 6, which includes also the
discusssion of the noise spectrum dependence on the parameters of the
detection coil and comparison of the results with those of other authors.

\section{Flux noise and current correlations}

In a flux noise experiment one measures and analyses the time
dependence of a voltage created in a detection coil by fluctuations of
currents in some conducting (or superconducting) object.
This voltage is determined by the time derivative of the magnetic flux
penetrating the coil, and the value  of the flux can be expressed
in terms of the current density distribution ${\bf j}({\bf r})$
[${\bf r}=(x_1,x_2,x_3)$] inside the object with the help of the
Biot-Savart's law, which in the Coulomb gauge ($\mbox{div}{\bf A}=0$)
can be written as
\begin{equation}
\Delta{\bf A}({\bf r})=-\mu_0{\bf j}^t({\bf r})\; , \label{lapla}
\end{equation}
where ${\bf A}({\bf r})$ is the vector potential defining the
distribution of the magnetic field (magnetic induction)
\mbox{${\bf B}({\bf r})=\mbox{rot}\,{\bf A}({\bf r})$}
created by ${\bf j}({\bf r})$,
\begin{equation}
\Delta\equiv
\frac{\partial^2}{\partial x_1^2}
+\frac{\partial^2}{\partial x_2^2}
+\frac{\partial^2}{\partial x_3^2}                      
\end{equation}
is the three-dimensional Laplacian and
\begin{equation}
{\bf j}^t({\bf r}) \equiv {\bf j}({\bf r})
-\Delta^{-1}\mbox{grad}\,\mbox{div}\,{\bf j}({\bf r})   \label{it}
\end{equation}
is the transverse part of ${\bf j}({\bf r})$.
Magnetic fields produced by the longitudinal part
of ${\bf j}({\bf r})$ cancel each other.

In the case of a system which can be considered as effectively
two-dimensional and situated (for simplicity) in the plane $x_3=0$ the
three-dimensional current density ${\bf j}({\bf r})$ is reduced to
\begin{equation}
{\bf j}({\bf r})={\bf i}({\bf x})\delta(x_3)\; ,          \label{i}
\end{equation}
where ${\bf x}\equiv x_\alpha$ ($\alpha=1,2$) is the two-dimensional
vector defining the position of a point in the plane   $x_3=0$ and
${\bf i }\equiv i_\alpha$ is the two-dimensional vector describing
the two-dimensional current density.

Substitution of Eq. (\ref{i}) into Eq. (\ref{lapla})
allows then to find that
\begin{eqnarray}
{\bf A}({\bf r}) & = & \mu_0\int \frac{d^2{\bf
q}}{(2\pi)^2}\int\frac{dq_3}{2\pi} \frac{\exp i({\bf
qx}+q_3x_3)}{q^2+q_3^2}{\bf i}^t({\bf q}) \nonumber
\\ & = & \frac{\mu_0}{2}\int \frac{d^2{\bf q}}{(2\pi)^2}
\frac{\exp(i{\bf qx}-q|x_3|)}{q}{\bf i}^t({\bf q})\;\, , \label{A}
\end{eqnarray}
where
\begin{equation}
{\bf i}({\bf q})=\int d^2{\bf x}\,
                 \exp(-i{\bf qx}) {\bf i}({\bf x})\,     \label{jq}
\end{equation}
is the (two-dimensional) Fourier transform of ${\bf i}({\bf x})$,
\mbox{$q=|{\bf q}|$} and
\begin{equation}
{\bf i}^t({\bf q})\equiv {\bf i}({\bf q}) -\hat{{\bf q}} ({\bf
\hat{q}i}({\bf q}))                  \label{jqt}
\end{equation}
is the transverse part of ${\bf i}({\bf q})$,
${\bf \hat{q}}\equiv{\bf q}/q$
being the unit vector parallel to ${\bf q}$.

In the simplest case a coil can be approximated by a closed circular
ring. Integration of ${\bf A}({\bf r})$ over the perimeter of the ring
$x_1^2+x_2^2=r^2$ situated at the distance $h$ from the plane $x_3=0$
gives
\begin{equation}
\Phi=\oint d{\bf r}\,{\bf A}({\bf r}) =\frac{\mu_0}{2}
\int\frac{d^2{\bf q}}{(2\pi)^2} F({\bf q})
i^t({\bf q})\; ,                                     \label{Phi}
\end{equation}
where
\begin{equation}
i^t({\bf q})=\sum_{\alpha,\beta}
\epsilon_{\alpha\beta}\hat{q}_\alpha i_\beta ({\bf q})
\end{equation}
is the amplitude of ${\bf i}^t({\bf q})$,
$\epsilon_{\alpha\beta}$ is the unit antisymmetric tensor,
\begin{equation}
F({\bf q})=\frac{2\pi r J_1(qr)}{q}\exp(-qh)        \label{D}
\end{equation}
is the geometrical factor depending on the parameters of the coil and
$J_1(z)$ is the first order Bessel function. In the case when the coil
can be considered as consisting of $N$ turns separated by the distance $b$
from each other, the expression for $F({\bf  q})$ should also include
an additional factor obtained by summation of contributions from
different turns \cite{JGRLM}:
\begin{equation}
F({\bf q})=\frac{2\pi r J_1(qr)}{q}\exp(-qh)
\frac{1-\exp(-Nqb)}{1-\exp(-qb)} \;\;.                 \label{gamma}
\end{equation}

The power spectrum of the flux noise is given by
the flux-flux correlation function
\begin{equation}
S(\omega)=\int dt \langle\Phi(t_0+t)\Phi(t_0)\rangle\exp(i\omega t)
                                                         \label{S}
\end{equation}
and with the help of Eq. (\ref{Phi}) can be expressed
in terms of the current density correlation function
\begin{eqnarray}
\langle{\bf i}^t {\bf i}^t\rangle_{{\bf q}\omega} \equiv
\int d^2{\bf x}\int & dt & \;\left[\exp(-i{\bf qx}+i\omega t)\right. \\
 & \times & \langle{\bf i}^t({\bf x_0+x},t_0+t)
\left.{\bf i}^t({\bf x_0},t_0)\rangle\right]    \nonumber
\end{eqnarray}
as
\begin{equation}
S(\omega)=\frac{\mu_0^2}{4} \int\frac{d^2{\bf q}}{(2\pi)^2}F^2({\bf q})
\langle{\bf i}^t {\bf i}^t\rangle_{{\bf q}\omega} \;\;. \label{S2}
\end{equation}

\section{The Hamiltonian of a Josephson junction array}

When self-induced magnetic field is taken into account
a square Josephson junction array can be described
by the Hamiltonian \cite{SK,DJ}
\begin{eqnarray}
H & = & -J\sum_{{\bf n},\alpha}\cos\left(\nabla_\alpha\varphi_{\bf n}
-A_{{\bf n}\alpha}\right)                 \nonumber \\ & &
+\frac{1}{2}\sum_{{\bf n,k}}(\nabla\times A)_{\bf n}M^{-1}_{\bf nk}
(\nabla\times A)_{\bf k} \;\;,                        \label{H}
\end{eqnarray}
where $\varphi_{\bf n}$ is the phase of the order parameter on the
${\bf n}$-th superconducting island, the variables $A_{{\bf n}\alpha}$
(defined on the bonds of the lattice) are determined
by the integral of the vector potential ${\bf A}({\bf r})$
over the line connecting the geometrical centers
of the neighboring superconducting islands:
\begin{equation}
A_{{\bf n}\alpha}=\frac{2e}{\hbar}
\int_{a{\bf n}}^{a({\bf n+e}_\alpha)}
d{\bf r}\,{\bf A}({\bf r})                         \label{Aja}
\end{equation}
and $a$ is period of the lattice.

The first term in Eq. (\ref{H})
describes the Josephson energy of the junctions in the array.
The coupling constant $J$ entering this term is determined by
the critical current $I_c$ of a single junction:
\begin{equation}
J=\frac{\hbar}{2e}I_c \;,                             \label{J}
\end{equation}
which is assumed to be the same for all junctions, whereas
$\nabla_\alpha\varphi_{\bf n}$
denotes the difference of $\varphi_{\bf n}$
between the neighboring sites of the lattice:
\begin{equation}
\nabla_\alpha\varphi_{\bf n}\equiv
\varphi_{{\bf n+e}_\alpha}-\varphi_{\bf n} \;.        \label{dphi}
\end{equation}
Here \mbox{${\bf e}_1=(1,0)$} and \mbox{${\bf e}_2=(0,1)$}.
Notice that the combination
\begin{equation}
\theta_{{\bf n}\alpha}\equiv\nabla_\alpha\varphi_{\bf n}
-A_{{\bf n}\alpha}                                    \label{th}
\end{equation}
which enters as the argument of the Josephson energy
$E_J(\theta)=-J\cos\theta$ is a gauge-invariant quantity.

The second term in Eq. (\ref{H}) is the energy of the magnetic field
\begin{equation}
E_{\rm mf}=\frac{1}{2\mu_0}\int d^3{\bf r}\,{\bf B}^2({\bf r})
\label{Emf} \end{equation}
expressed in terms of the variables $A_{{\bf n}\alpha}$.
The matrix $M_{\bf nk}\equiv M({\bf n-k})$ is usually called
the mutual inductance matrix \cite{SK,DJ,PZWO} and
\begin{equation}
(\nabla\times A)_{\bf n}\equiv\sum_{\alpha,\beta}
\epsilon_{\beta\alpha}\nabla_\beta A_{{\bf n}\alpha}   \label{rotA}
\end{equation}
is the directed sum of the variables $A_{{\bf j}\alpha}$
along the perimeter of a lattice plaquette
(the lattice equivalent of $\mbox{rot}\,{\bf A}$)
and is proportional to the magnetic flux penetrating the plaquette.

Variation of Eq. (\ref{H}) with respect to $A_{{\bf n}\alpha}$
gives the equation
\begin{equation}
I_{{\bf n}\alpha}=\frac{2e}{\hbar}\sum_{{\bf k}}\widetilde{\nabla}
                  \times M^{-1}_{\bf nk}(\nabla\times A)_{\bf k}
                  \;,                                  \label{I1}
\end{equation}
which relates the value of the superconducting current through a
junction \begin{equation}
I_{{\bf n}\alpha}=I_c\sin(\nabla_\alpha\varphi_{\bf n}-A_{{\bf
n}\alpha})                                           \label{I}
\end{equation}
with the vector potential of the magnetic field induced
by the presence of the currents in the array.
Here [like in Eq. (\ref{rotA})] $\widetilde{\nabla}\times$ stands for
$\sum_{\beta}\epsilon_{\beta\alpha}\widetilde{\nabla}_\beta$, whereas
$\widetilde{\nabla}_\beta$ designates the lattice difference,
analogous to the one defined by Eq. (\ref{dphi}), but shifted
in the negative direction:
\begin{equation}
\widetilde{\nabla}_\beta X_{\bf n}\equiv X_{\bf n}-X_{{\bf n-e}_\beta}
\;.                                                 \label{nablat}
\end{equation}

On the other hand variation  of Eq. (\ref{H}) with respect
to $\varphi_{\bf n}$ gives the current conservation equation
\begin{equation}
(\widetilde{\nabla}I)_{{\bf n}}=0 \;,            \label{CC}
\end{equation}
where
\begin{equation}
(\widetilde{\nabla}I)_{{\bf n}}\equiv
\sum_{\alpha}[I_{{\bf n}\alpha}-I_{({\bf n-e}_\alpha)\alpha}]
                                                      \label{divI}
\end{equation}
is the lattice equivalent of divergence.
Eq. (\ref{CC}) can be alternatively obtained by application of the
operator $\widetilde{\nabla}_\alpha$ to Eq. (\ref{I1}).
Therefore Eq. (\ref{I1}) and Eq. (\ref{CC})
[both obtained by variation of Eq. (\ref{H})]
are not independent of each other.

The vector potential of the magnetic field created by the currents
flowing in the array
can be chosen purely transverse ($\mbox{div}{\bf A}=0$), which in
terms of $A_{{\bf n}\alpha}$ corresponds to
\begin{equation}
\widetilde{\nabla}_\alpha A_{{\bf n}\alpha}=0 \;.         \label{divA}
\end{equation}
In that case Eq. (\ref{I1}) is reduced to
\begin{equation}
I_{{\bf n}\alpha}=\frac{2e}{\hbar}\sum_{{\bf k}} M^{-1}_{\bf nk}
                  (-\Delta_L A_\alpha)_{\bf k} \;,       \label{I2}
\end{equation}
where
$\Delta_L\equiv\sum_{\beta}\widetilde{\nabla}_\beta\nabla_\beta$
is the two-dimensional lattice analog of the Laplacian:
\begin{equation}
(\Delta_L X)_{\bf k}=\sum_{\beta}
(X_{{\bf k+e}_\beta}-2X_{\bf k}+X_{{\bf k-e}_\beta}) \;. \label{lapl}
\end{equation}
Comparison of Eq. (\ref{I2}) with Eq. (\ref{A})
allows to find that for $|{\bf n}-{\bf k}|\gg 1$
\begin{equation}
M^{-1}_{\bf nk}\approx\left(\frac{\hbar}{2e}\right)^2
\frac{1}{\pi\mu_0  a|{\bf n}-{\bf k}|} \;,           \label{M}
\end{equation}
whereas for $|{\bf n}-{\bf k}|\sim 1$ the form of $M^{-1}_{{\bf nk}}$
depends on the particular shape of superconducting islands \cite{PZWO}.

Linearization of Eqs. (\ref{I1}) and their solution allows to show that
when the magnetic fields of the currents in the array are taken into
account, the logarithmic interaction of vortices becomes
screened \cite{SK} at so-called magnetic field penetration length
$\Lambda$, exactly as it happens in superconducting films \cite{Pearl}.
When screening is relatively weak (that is when $\Lambda\gg a$),
the value of $\Lambda$ is given by
\begin{equation}
\Lambda\approx \frac{2}{\mu_0 J}\left(\frac{\hbar}{2e}\right)^2
                                                   \label{Lambda-1}
\end{equation}
and does not depend on the shape of superconducting islands forming
the array \cite{SK}.

Instead of considering Hamiltonian (\ref{H}) as dependent on two
different types of variables defined on the sites ($\varphi_{\bf n}$)
and  on the bonds ($A_{{\bf n}\alpha}$) of the lattice, it is
convenient to use a single variable,
namely the gauge invariant phase difference
$\theta_{{\bf n}\alpha}$ defined by Eq. (\ref{th}).
In terms of $\theta_{{\bf n}\alpha}$ the Hamiltonian (\ref{H})
can be rewritten as
\begin{equation}
H=-J\sum_{{\bf n},\alpha}\cos\theta_{{\bf n}\alpha}
+\frac{1}{2}\sum_{{\bf n,k}}(\nabla\times \theta)_{\bf n}M^{-1}_{\bf
nk} (\nabla\times \theta)_{\bf k} \;,                 \label{Hth}
\end{equation}
variation of which with respect to $\theta_{{\bf n}\alpha}$
reproduces Eq. (\ref{I1}) in the form
\begin{equation}
I_{{\bf n}\alpha}=-\frac{2e}{\hbar}\sum_{{\bf k}}\widetilde{\nabla}
                  \times M^{-1}_{\bf nk}(\nabla\times \theta)_{\bf k}
                  \;,                                 \label{I1th}
\end{equation}
where the expression for the superconducting current
\begin{equation}
I_{{\bf n}\alpha}=I_c\sin\theta_{{\bf n}\alpha}         \label{Ith}
\end{equation}
is naturally consistent with Eq. (\ref{I}).
As previously, the current conservation equation (\ref{CC})
can be obtained by application of the operator
$\widetilde{\nabla}_\alpha$ to Eq. (\ref{I1th}).

\section{Dynamic fluctuations in array of resistively shunted
junctions}

The dynamic description of the same system requires to complement
the Hamiltonian $H$ by the dissipative function $W$
(we assume that the array is overdamped and therefore its dynamics
is purely relaxational).
In the case of the array formed by SNS (superconductor - normal metal -
superconductor) junctions one can describe dissipation in terms of
the effective resistance shunting each junction (so-called RSJ-model).
This corresponds to $W\{\theta_{}\}$ of the form
\begin{equation}
W=\eta\sum_{{\bf n},\alpha}\left(\frac{\partial}{\partial  t}
\theta_{{\bf n}\alpha}\right)^2,                       \label{W}
\end{equation}
where the effective viscosity
\begin{equation}
\eta=\left(\frac{\hbar}{2e}\right)^2\frac{1}{R}       \label{eta}
\end{equation}
is determined by the value of the shunting resistance $R$,
which is assumed to be the same for all junctions.
For $W$ of the form (\ref{W}) the conservation of
energy is achieved when the time evolution of the variables
$\theta_{{\bf n}\alpha}$ is governed by the standard equations of
relaxational dynamics:
\begin{equation}
\eta \frac{\partial}{\partial  t}\theta_{{\bf n}\alpha}
=-\frac{\partial H}{\partial  \theta_{{\bf n}\alpha}} \;.\label{Lang}
\end{equation}

On the other hand, Eq. (\ref{Lang}) can be rewritten in the form
(\ref{I1th}), where the expression for the current should be replaced
by
\begin{equation}
I_{{\bf n}\alpha}=I_c\sin\theta_{{\bf n}\alpha}
+\frac{\hbar}{2eR}\frac{\partial}{\partial  t} \theta_{{\bf n}\alpha}
\;.                                                   \label{ISN}
\end{equation}
The time derivative of $\theta_{{\bf n}\alpha}$ being proportional
to the voltage,
the second term in Eq. (\ref{ISN}) can be easily identified as the
normal current flowing through the junction.
Consideration of purely relaxational dynamics means that we are
neglecting capacitive effects  and currents have to be conserved on
each site of the lattice (in other words, only transverse current are
allowed). This is ensured by the form of Eq.
(\ref{I1th}), substitution of which into Eq. (\ref{CC}) automatically
leads to its fulfillment for any form of $I_{{\bf n}\alpha}$.

In presence of thermal fluctuations the right-hand side of
Eq. (\ref{Lang}) should be complemented with
the random force term $\xi_{{\bf n}\alpha}(t)$:
\begin{equation}
\eta \frac{\partial}{\partial  t}\theta_{{\bf n}\alpha}
=-\frac{\partial H}{\partial  \theta_{{\bf n}\alpha}}
 +\xi_{{\bf n}\alpha}+f_{{\bf n}\alpha} \;,      \label{Lang2}
\end{equation}
the correlations of which are Gaussian and satisfy
\begin{equation}
\langle\xi_{{\bf n}\alpha}(t)\xi_{{\bf k}\beta}(t')\rangle
=2\eta T\delta_{{\bf nk}}\delta_{\alpha\beta}\delta(t-t') \label{xi}
\;, \end{equation}
where $T$ is the temperature expressed in energy units (that is
multiplied by the Boltzmann constant $k_B$).
We also have included in the right-hand side of Eq. (\ref{Lang2}) the
non-random external force $f_{{\bf n}\alpha}$ (to be discussed later).

In terms of the expression for the current the inclusion into Eq.
(\ref{Lang2}) of the random term $\xi_{{\bf n}\alpha}$ corresponds
to appearance in Eq. (\ref{ISN}) of the additional (fluctuating)
contribution to normal current $\delta I_{{\bf n}\alpha}$:
\begin{equation}
I_{{\bf n}\alpha}=I_c\sin\theta_{{\bf n}\alpha}
+\frac{\hbar}{2eR} \frac{\partial}{\partial  t} \theta_{{\bf n}\alpha}
+\delta I_{{\bf n}\alpha} \;,                        \label{ISNF}
\end{equation}
where $\delta I_{{\bf n}\alpha}\equiv -(2e/\hbar)\xi_{{\bf n}\alpha}$.
Note that since Eq. (\ref{I1th}) describes the relation between
the currents and the magnetic field induced by them, it has to remain
fulfilled also when fluctuations of currents are taken
into account. The validity of the current conservation equations
(\ref{CC}) remains ensured
by the form of the right-hand side of Eq. (\ref{I1th}).

The suggestion to describe the dynamics of a resistively shunted
Josephson junction array by Eqs. (\ref{ISNF}) has been put forward by
Shenoy \cite{Shenoy}, who did not consider fluctuations of the
magnetic field, that is assumed \makebox{$\theta_{{\bf n}\alpha}\equiv
\varphi_{{\bf n}+{\bf e}_\alpha}-\varphi_{\bf n}$}.
In that case substitution of Eqs. (\ref{ISNF}) into current
conservation equations (\ref{CC}) leads to dynamic equations for
$\varphi_{\bf n}$ with non-local effective viscosity \cite{Shenoy}.
Quite remarkably the inclusion into consideration
of the magnetic field fluctuations
leads to simplification of the dynamic equations which (in terms of
$\theta_{{\bf n}\alpha}$) become local.
The idea that in presence of magnetic field fluctuations
a resistively shunted Josephson junction array can be
described by Eqs. (\ref{Lang2}), where $\theta_{{\bf n}\alpha}$ is
the gauge invariant phase difference, has been introduced by
Dom\'{i}nguez and Jos\'{e} \cite{DJ}.

\section{The fluctuation-dissipation theorem}

It is well known that when the time evolution  of some variables
$\{\theta\}$ is governed by the standard Langevin equations of
the form (\ref{Lang2}),
the equilibrium (that is calculated for $f_{{\bf n}\alpha}= 0$)
correlation function
\begin{equation}
C_{{\bf n}\alpha,{\bf k}\beta}(t-t')\equiv
\langle\theta_{{\bf n}\alpha}(t)\theta_{{\bf k}\beta}(t')\rangle
_{f=0}                                          \label{cth}
\end{equation}
is related with the response function
\begin{equation}
G_{{\bf n}\alpha,{\bf k}\beta}(t-t')
\equiv\left.\frac{\delta \langle \theta_{{\bf n}\alpha}\rangle}
{\delta f_{{\bf k}\beta}} \right|_{f=0}            \label{g}
\end{equation}
by the fluctuation-dissipation theorem:
\begin{equation}
G_{{\bf n}\alpha,{\bf k}\beta}(t)
-G_{{\bf n}\alpha,{\bf k}\beta}(-t)
=-\frac{1}{T}\frac{\partial}{\partial t}
C_{{\bf n}\alpha,{\bf k}\beta}(t) \; .    \label{fdt}
\end{equation}

However, in practical situation one is interested not in the
response of the gauge invariant phase difference $\theta_{{\bf
n}\alpha}$ to the (unspecified) conjugate force $f_{{\bf n}\alpha}$,
but rather in  more readily observable quantities such as
conductivity, which is the response of a current to application
of electric field. In situation when electric field ${\bf E(r)}$
is created due to presence of ac magnetic field it is given by
\begin{equation}
{\bf E}({\bf r})=
-\frac{\partial}{\partial t}{\bf A}({\bf r}) \; , \label{E}
\end{equation}
where ${\bf A}({\bf r})$ is the vector potential defining the
(total) magnetic field ${\bf B}({\bf r})=\mbox{rot}\,{\bf A}({\bf
r})$.

In presence of the external magnetic field
\mbox{${\bf B}^{\rm e}({\bf r})=\mbox{rot}\,{\bf A}^e({\bf r})$}
the expression (\ref{Emf}) describing the magnetic field energy
should be replaced by the expression for the Gibbs free energy
\begin{equation}
F_{\rm mf}=\frac{1}{2\mu_0}\int d^3{\bf r}[{\bf B}^2({\bf r})
          -2{\bf B(r)B^{\rm e}(r)}] \;,               \label{Emf2}
\end{equation}
variation of which in absence of any other terms gives
${\bf B(r)}={\bf B}^{\rm e}({\bf r})$. This leads to appearance in
the Hamiltonian (\ref{Hth}), describing the array, of the additional term
\begin{equation}
\Delta H=\sum_{{\bf n,k}}(\nabla\times\theta)_{{\bf n}}
M^{-1}_{{\bf n}{\bf k}}(\nabla\times A^{\rm e})_{{\bf k}} \;\; . \label{dH}
\end{equation}
Here the variables
\begin{equation}
A^{\rm e}_{{\bf n}\alpha}
=\frac{2e}{\hbar}\int_{a{\bf n}}^{a({\bf n+e}_\alpha)}
d{\bf r}\,{\bf A}^e({\bf r}) \;,                      \label{Ajae}
\end{equation}
defined on lattice bonds, are related to the vector potential
${\bf A}^{\rm e}({\bf r})$ of the external magnetic field exactly in
the same way as earlier introduced variables $A_{{\bf n}\alpha}$
are related to the total vector potential ${\bf A}({\bf r})$.

The form of the correction to the Hamiltonian given by
Eq. (\ref{dH}) corresponds to the presence in Eq.
(\ref{Lang2}) of the external force term
\begin{equation}
f_{{\bf n}\alpha}=-\sum_{{\bf k}}\widetilde{\nabla}
     \times M^{-1}_{\bf nk}(\nabla\times A^{\rm e})_{\bf k} \;.
                                                      \label{f}
\end{equation}
Comparison of Eq. (\ref{f}) with Eq. (\ref{I1th}) shows that
(up to the factor of $2e/\hbar$)
$f_{{\bf n}\alpha}$ is related to $A^{\rm e}_{{\bf n}\alpha}$
exactly in the same way as
$I_{{\bf n}\alpha}$ is related to $\theta_{{\bf n}\alpha}$.
This allows to conclude that
the correlation functions of the currents in the array
\begin{equation}
C^I_{{\bf n}\alpha,{\bf k}\beta}(t-t')\equiv
\langle I_{{\bf n}\alpha}(t)I_{{\bf k}\beta}(t')\rangle
                                                        \label{ci}
\end{equation}
and the response function
\begin{equation}
G^I_{{\bf n}\alpha,{\bf k}\beta}(t-t')
\equiv\left.\frac{\delta\langle I_{{\bf n}\alpha}\rangle}
{\delta (A^{\rm e}_{{\bf k}\beta})^t} \right|_{f=0}     \label{gi}
\end{equation}
have to satisfy the relation
\begin{equation}
G^I_{{\bf n}\alpha,{\bf k}\beta}(t)
-G^I_{{\bf n}\alpha,{\bf k}\beta}(-t)
=-\frac{\hbar}{2e}\frac{1}{T}\frac{\partial}{\partial t}
C^I_{{\bf n}\alpha,{\bf k}\beta}(t)                  \label{fdt2}
\end{equation}
completely analogous to Eq. (\ref{fdt}). Here $(A^{\rm e}_{{\bf
k}\beta})^t$ is the transverse part of $A^{\rm e}_{{\bf k}\beta}$.
As can be seen from the right-hand side of Eq. (\ref{f}),
the longitudinal part of $A^{\rm e}_{{\bf k}\beta}$ is completely
decoupled from fluctuations of $\theta_{{\bf n}\alpha}$.

In terms of effective conductivity
$g^{\rm eff}_{{\bf n}\alpha,{\bf k}\beta}$,
defined as the response function
\begin{equation}
g^{\rm eff}_{{\bf n}\alpha,{\bf k}\beta}(t-t')
\equiv\frac{\delta}{\delta V^{\rm e}
_{{\bf k}\beta}} \left. \langle I_{{\bf
n}\alpha}\rangle\right|_{V^{\rm e}=0}               \label{geff}
\end{equation}
of the current with respect to external voltage
\begin{equation}
V^e_{{\bf k}\beta}=-\frac{\hbar}{2e}\frac{\partial}{\partial t}
(A^{\rm e}_{{\bf k}\beta})^t\;,                      \label{v}
\end{equation}
Eq. (\ref{fdt2}) can be rewritten as
\begin{equation}
C^I_{{\bf n}\alpha,{\bf k}\beta}(t)                  \label{fdt3}
=T[g^{\rm eff}_{{\bf n}\alpha,{\bf k}\beta}(t)
+g^{\rm eff}_{{\bf n}\alpha,{\bf k}\beta}(-t)] \;.
\end{equation}

For wavevectors small in comparison with $a^{-1}$, the variables $I_{{\bf
n}\alpha}$ can be identified with $a{\bf i}({\bf x})$, whereas
$V^{\rm e}_{{\bf n}\alpha}$ with $a{\bf E}^e_\parallel({\bf x})$, where
${\bf E}^{\rm e}_\parallel({\bf x})$ is the projection of the external
electric field ${\bf E}^{\rm e}({\bf x})$ on the plane $x_3=0$ (here and
below we assume that ${\bf E}^{\rm e}_\parallel({\bf q},\omega)$ is
transverse). This allows to rewrite Eq. (\ref{fdt3})  as
\begin{equation}
\langle {\bf i}^t{\bf i}^t\rangle_{{\bf q}\omega}=
2T\mbox{Re}\left[{g^{\rm eff}({\bf q},\omega)}\right] \;, \label{fdt4}
\end{equation}
where the Fourier transform $g^{\rm eff}({\bf q},\omega)$
of the effective conductivity is the coefficient of proportionality
in the relation
\begin{equation}
{\bf i}({\bf q},\omega)
=g^{\rm eff}({\bf q},\omega)
{\bf E}^{\rm e}_\parallel({\bf q},\omega) \;.             \label{geff2}
\end{equation}

One should distinguish between $g^{\rm eff}({\bf q},\omega)$
and (also momentum and frequency dependent)
sheet conductance $g_\Box({\bf q},\omega)$,
which is defined as the coefficient of proportionality between
${\bf i}({\bf q},\omega)$ and {\em total} electric field ${\bf
E}_\parallel({\bf q},\omega)$:
\begin{equation}
{\bf i}({\bf q},\omega)
=g_\Box{\bf q},\omega)
{\bf E}_\parallel({\bf q},\omega) \;.               \label{zsh}
\end{equation}
The form of  the current induced correction to
${\bf E}^{\rm e}_\parallel({\bf q},\omega)$ can be easily found with
the help of Eq. (\ref{A}) and Eq. (\ref{E}), which lead to
\begin{equation}
{\bf E}_\parallel({\bf q},\omega)
={\bf E}^{\rm e}_\parallel({\bf q},\omega)
+i\omega\frac{\mu_0 }{2q}{\bf i}({\bf q},\omega) \;. \label{etotal}
\end{equation}
Substitution of 
Eq. (\ref{zsh}) into Eq. (\ref{etotal}) then gives
\begin{equation}
{\bf E}_\parallel({\bf q},\omega) =\frac{1}{1-i\omega\frac{\mu_0 }{2q}
g_\Box({\bf q},\omega)} {\bf E}^{\rm e}_\parallel({\bf q},\omega)
                                                   \label{etotal2}
\end{equation}
and, accordingly,
\begin{equation}
g^{\rm eff}({\bf q},\omega)=\frac{1}{- i\omega\frac{\mu_0}{2q}
+Z_\Box({\bf q},\omega)} \;\;,                     \label{zeff3}
\end{equation}
where $Z_{\Box}({\bf q},\omega)\equiv 1/g_{\Box}({\bf q},\omega)$
is momentum and frequency dependent sheet impedance.

The form of Eq. (\ref{zeff3}) suggests that the response of a current
in a two-dimensional system to external electric field is the same as
if the proper sheet impedance of a system $Z_{\Box}({\bf q},\omega)$
has been connected in series with the other contribution,
which can be considered as the
effective impedance of the empty space surrounding this system.
This additional contribution is purely inductive and corresponds to
\begin{equation}
L_s(q)=\frac{\mu_0}{2q} \;.                          \label{L0}
\end{equation}

In the case of a superconducting system in the low frequency limit
$Z_\Box({\bf q},\omega)\approx -i\omega L_\Box$, where $L_\Box$ is the
effective sheet inductance, substitution of which into
Eq. (\ref{etotal2}) allows to rewrite it as
\begin{equation}
{\bf E}_\parallel({\bf q},\omega)
=\frac{\Lambda q}{1+\Lambda q}
{\bf E}^{\rm e}_\parallel({\bf q},\omega)         \label{etotal3}
\end{equation}
where
\begin{equation}
\Lambda=\frac{2L_\Box}{\mu_0}
\label{Lambda} \end{equation}
is the (two-dimensional) magnetic field penetration length \cite{Pearl}
already discussed in Sec. 3.
Since in the considered system electric field appears only due
to presence of ac magnetic field, the same length describes as well
the screening of the electric field.

The form of the fluctuation-dissipation theorem obtained after
substitution of Eq. (\ref{zeff3}) into Eq. (\ref{fdt4})
\begin{equation}
\langle{\bf i}^t{\bf i}^t\rangle_{{\bf q}\omega}
=2T\mbox{Re}\frac{1}{-i\omega L_s(q)+Z_\Box({q},\omega)} \label{fdt5}
\end{equation}
being completely independent of the details of the structure of
the particular system used for its derivation,
one can expect it also to be valid for other two-dimensional
superconducting (or simply conducting) systems, in particular
superconducting films.

\section{Results and discussion}

Substitution of Eq. (\ref{fdt5}) into Eq. (\ref{S2}) gives
the expression for the flux noise power spectrum
\begin{equation}
S(\omega)=\frac{\mu_0^2 T}{2} \int\frac{d^2{\bf q}}{(2\pi)^2}F^2(q)
\mbox{Re}\left[\frac{1}{-i\omega L_s(q)+Z_\Box(q,\omega)}\right]
                                                      \label{S3a}
\end{equation}
which is the central result of this work.
It allows instead of constructing special theories explaining
frequency dependence of $S(\omega)$ in different regimes
to relate it with already known properties of $Z_\Box(q,\omega)$.

It is of interest to compare Eq. (\ref{S3a}) with the expression for
the quantity
[the correction to frequency dependent mutual impedance $Z_m(\omega)$],
which is measured in the framework of the two coil method and defined
as the ratio of the measured voltage to the driving current.
In our notation this expression \cite{JGRLM} can be rewritten as
\begin{equation} \delta
Z_m(\omega)=-\omega^2\frac{\mu_0^2 }{4} \int\frac{d^2{\bf q}}{(2\pi)^2}
\frac{{F}(q)\widetilde{F}(q)}{-i\omega L_s(q)+Z_\Box(q,\omega)} \;,
                                                      \label{Zm}
\end{equation}
where $\widetilde{F}(q)$ is the geometrical factor analogous to $F(q)$,
but describing the driving coil.
Note that applicability of Eq. (\ref{S3a}) and Eq. (\ref{Zm}) for
description of a Josephson junction array requires the lattice
constant of the array to be much smaller then the geometrical parameters
of the coil(s) ($r$ and $h$) entering the expressions for $F(q)$
[and $\widetilde{F}(q)$].

Comparison of Eq. (\ref{Zm}) with Eq. (\ref{S3a}) shows that
the real part of $\delta Z_m(\omega)$ is determined by
the same (real) component of 
$g^{\rm eff}(q,\omega)$
as the noise spectrum, so in absence of a difference between $F(q)$ and
$\widetilde{F}(q)$ the noise spectrum $S(\omega)$ would be simply
proportional to $\mbox{Re}[\delta Z_m(\omega)]$:
\begin{equation}
S(\omega)=-\frac{2T}{\omega^2}\mbox{Re}[\delta Z_m(\omega)] \;. \label{S3b}
\end{equation}

In the case of the simplest coil
(a single circular loop of radius $r$) Eq. (\ref{S3a}) is reduced to
\begin{eqnarray}
S(\omega)={\pi}\mu_0^2Tr^2\int_{0}^{\infty} & dq & \;\left\{
\frac{J^2_1(rq)}{q}\exp(-2hq)\right. \label{S3} \\
 & \times & \left.\mbox{Re}\left[\frac{1}{-i\omega
L_s(q)+Z_\Box(q,\omega)}\right]\right\}          \nonumber
\end{eqnarray}
The analogous equation derived by Kim and Minnhagen \cite{KM}
differs from Eq.
(\ref{S3}) basically (i) by the absence of the term $L_s(q)$ and (ii)
by the presence in the integrand of the additional factor $q^4$.
The former of the two discrepancies is rather natural,
since in Ref. \onlinecite{KM} the screening effects
have not been taken into account,
whereas the latter we believe to be the consequence of the incorrect
calculation of the magnetic field produced by currents
in the array.

In Ref. \onlinecite{KM} this magnetic field is calculated as the sum of
magnetic fields produced by current loops  associated with lattice
plaquettes, whereas the magnitude of a current in each loop is assumed
to be given the directed sum of the currents in the junctions $I_{{\bf
n}\alpha}$ along the perimeter of a plaquette (in the present work
this sum is denoted $\nabla\times I$). In such procedure the current of
each junction is counted twice (as giving contributions to the loop
currents associated with the two neighboring lattice plaquettes) but
with opposite signs, so these two contributions almost cancel each other,
which leads to appearance of the additional $q$-dependent factor in
comparison with our Eq. (\ref{S3}). In a more consistent implementation
of this approach the values of loop currents associated with arrays
plaquettes should be chosen in such a way that the value of the current
on each junction is given by the difference of the loop currents
associated with the two neighboring plaquettes. These  so called mesh
currents \cite{PZWO,DK} $I^{\rm m}_{\bf n}$ are related with the
currents on the junctions $I_{{\bf n}\alpha}$ as
\begin{equation}
I^{\rm m}_{\bf n}\equiv -\Delta_L^{-1}(\nabla\times I)_{\bf n} \label{Im}
\end{equation}
which explains the appearance of the additional factor of $q^4$ in
calculation which uses $(\nabla\times I)_{\bf n}$ instead of
$I^{\rm m}_{\bf n}$.

The main contribution to the integral in Eq. (\ref{S3}) is coming from
the region
\begin{equation}
0<q\lesssim 1/h,                                         \label{qh}
\end{equation}
so in the situation when $h$ exceeds all the microscopic scales
responsible for the $q$ dependence of $Z_\Box(q,\omega)$, one can
replace in Eq. (\ref{S3}) $Z_\Box(q,\omega)$ by
\begin{equation}
Z_\Box(\omega)
\equiv-i\omega L_\Box(\omega)+R_\Box(\omega)   \label{Zomega}
=\lim_{q\rightarrow 0}Z_\Box(q,\omega)
\end{equation}
and use $S(\omega)$ to extract information about $L_\Box(\omega)$ and
$R_\Box(\omega)$.

Let us first discuss the limit when the effects of screening can be
neglected. This is possible if in the essential part of the interval
(\ref{qh}) one can neglect $L_s(q)$ in comparison with $L_\Box(\omega)$
and therefore requires $h\ll\Lambda$ (and not $h\gg\Lambda$ as
has been claimed in Ref. \onlinecite{Festin} and Ref. \onlinecite{KM}).
In that case Eq. (\ref{S3}) is reduced to
\begin{equation}
S(\omega)=\pi\mu_0^2 T r^2
Y\left({h}/{r}\right)\frac{R_\Box(\omega)} {\omega^2
L_\Box^2(\omega)+R_\Box^2(\omega)} \;,        \label{S4}
\end{equation}
where
\begin{eqnarray}
Y\left({h}/{r}\right) & = &
\int_{0}^{\infty}dq\,\frac{J_1^2(rq)}{q}\exp\left(-2hq \right)
                                                     \label{Y} \\
 & \approx & \left\{\begin{array}{ll}
\frac{1}{2} & \mbox {for } h/r\ll 1 \\
\frac{1}{16}\left(\frac{r}{h}\right)^2 & \mbox{for } h/r\gg 1 \;\;.
\end{array}\right. \nonumber
\end{eqnarray}
This means that with increase of $h$ the amplitude of the noise has
first to change very slowly and then to decay proportionally to
$1/h^2$. The experimental results of Festin {\em et al.} \cite{Festin},
obtained on thin YBCO film at $h/r\sim 1$,
are compatible with $1/h^2$
dependence even better than
with $1/h^3$ dependence predicted by Kim and Minnhagen \cite{KM}.

In the opposite limit of strong screening ($\Lambda\ll h$)
$L_\Box(\omega)$ is negligibly small in comparison with $L_s(\omega)$
and $S(\omega)$ depends only on $R_\Box(\omega)$, but not on
$L_\Box(\omega)$. In particular, for $h\ll r$ integration
in Eq. (\ref{S3}) gives
\begin{equation}
S(\omega)\approx\left\{\begin{array}{ll} \frac{\pi\mu_0 rT}{\omega} &
\mbox{for } \omega\ll \frac{R_\Box(\omega)}{\mu_0h} \\
\frac{2r}{h}\frac{R_\Box(\omega)T}{\omega^2} & \mbox{for } \omega\gg
\frac{R_\Box(\omega)}{\mu_0h} \;\;.
\end{array}\right.                              \label{S5}
\end{equation}
Note that for $\Lambda\ll h\ll r$ and small $\omega$ one obtains
the dependence $S(\omega)\propto 1/\omega$, which is observed in noise
measurements in Josephson junction arrays \cite{Shaw,Candi} over four
decades in frequency.

The idea that this particular dependence appears because
SQIUD integrates contributions from wide interval of wavevectors has
been put forward by Wagenblast and Fazio \cite{WF}. Our analysis has
shown that in order to obtain the $1/\omega$ dependence over four
decades in frequency one should have $r/h \gtrsim 10^4$, which
in the experiments of Refs. \onlinecite{Shaw,Candi}
definitely has not been fulfilled.
Thus the origin of the $1/\omega$ dependence of the flux noise
power spectrum observed experimentally in Josephson junction arrays
(in particular, at tempratures lower then the expected temperature
of the Berezinskii-Kosterlitz-Thouless (BKT) phase transition
\cite{Ber,KT}) still remains to be elucidated.

Substitution of the results of Ambegaokar {\em et al.} \cite{AHNS}
for vortex pair contribution to impedance into Eq. (\ref{S3a})
or its simplified analogs allows to reproduce the dependence
$S(\omega)\propto 1/\omega$ only exactly at the temperature of the BKT
transition. On the other hand,
the analogous frequency dependence of the noise spectrum
observed in thin YBCO films \cite{Ferrari} is
ascribed to vortex hopping between neighboring pinning centers,
a finite vortex concentration being associated with presence
of a residual magnetic field. The same mechanism may be responsible
for the results of experiments \cite{Lerch,Shaw,Candi}
on Josephson junction arrays.

\section*{Acknowledgments}

The author is grateful to V. V. Lebedev for useful discussion.
This work has been supported by the Program
"Quantum Macrophysics" of the Russian Academy of Sciences,
by the Program "Scientific Schools of the Russian Federation" (grant No.
00-15-96747), by the Swiss National Science Foundation,
and by the Netherlands Organization for Scientific Research (NWO)
in the framework of Russian-Dutch Cooperation Program.

\end{multicols}

\end{document}